\documentclass{iaus}

\usepackage{amsmath}
\usepackage{amsfonts}
\usepackage{amssymb}
\usepackage{graphicx}
\usepackage{float}

\title{Revised element abundances for WC-type central stars}
\author{Helge Todt, G\"otz Gr\"afener, Wolf-Rainer Hamann}
\affiliation{Institut f\"ur Physik, Universit\"at Potsdam, 
Am Neuen Palais 10, 14469 Potsdam, Germany 
\break email: htodt@astro.physik.uni-potsdam.de}

\pubyear{2006}
\volume{234}  
\pagerange{127--130}
\jname{Planetary Nebulae in Our Galaxy and Beyond}
\editors{M. J. Barlow \& R. H. M\'endez, eds.}
\doi{10.1017/S1743921306002869}

\begin{document}

\maketitle

\begin{abstract}According to previous spectral analyses of Wolf-Rayet
type  central stars, late [WC] subtypes show systematically higher
carbon-to-helium  abundance ratios than early [WC] subtypes. If this
were true, it would rule out that these stars form an evolutionary
sequence. However, due to the different parameter domains and
diagnostic lines, one might suspect systematic errors being the source
of this discrepancy. In an ongoing project we are therefore checking
the [WC] analyses by means of the last generation of non-LTE models for
expanding stellar atmospheres  which account for line-blanketing and
wind clumping. So far, the abundance discrepancy is not resolved.
Further element abundances (H, N, Fe) are determined and compared
with evolutionary predictions. 
\end{abstract}


\section{Introduction}

Most central stars of planetary nebulae have a hydrogen-rich
atmosphere; their optical spectra show only weak absorption lines,
mainly of hydrogen and helium. In drastic contrast, there is another
class of central stars whose spectra are characterized by  strong
emission lines of carbon, helium and oxygen. Very much like in the case
of the Population\,I Wolf-Rayet stars, these spectra emerge from a fast
and dense stellar wind. The spectral type of these hydrogen-deficient
central stars is denoted as [WC], where  the C indicates the dominance
of carbon lines, while the brackets should distinguish them from their
massive counterparts.  

The [WC] class is divided into a sequence of subtypes, from [WC2] to
[WC11] (a more recent scheme introduces also subtypes [WO1-4]). 
The subtypes [WC2-5] are referred to as ``early'' types or
[WCE], while [WC6-11] subtypes are considered as ``late'', briefly
[WCL]. [WCE] types show lines of C\,{\sc{iv}}, He\,{\sc{ii}} and
oxygen, whereas late-type spectra are dominated by C\,{\sc{ii}},
C\,{\sc{iii}} and C\,{\sc{iv}}. Hence, the sequence
of decreasing subtype number corresponds to increasing ionization, and
therefore increasing effective temperature. If central stars evolve
directly from the AGB to the white dwarf stage, they should first become
[WCL] and then [WCE] subtypes.

\section{Hydrogen deficiency}

\begin{figure}[t]
\begin{center}
\includegraphics{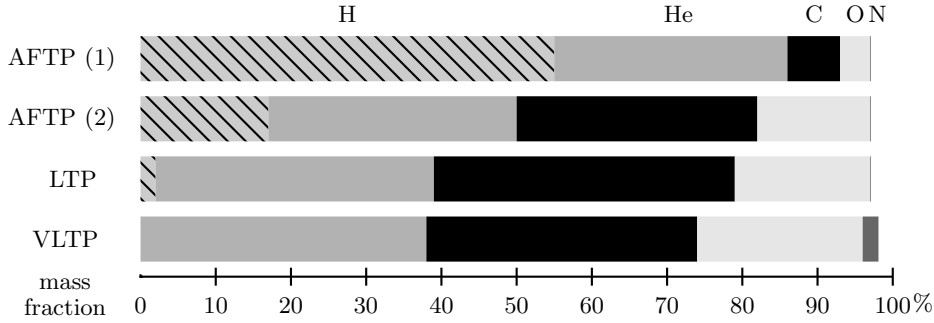}
\caption{Stellar surface abundances predicted from stellar evolution models
with simultaneous burning and mixing (Herwig \cite{Herwig01}, 
Werner \& Herwig \cite{Werner06}). The predictions depend on the adopted 
scenario for the moment of the last thermal pulse; two different models 
are given for the AFTP case, depending on the H envelope mass.} 
\label{fig:LTPabundances}
\end{center}
\end{figure}

It is generally adopted that [WC]-type central stars have lost their
hydrogen envelope in a last thermal pulse following the AGB evolution.
However, it is not clear whether this last thermal pulse occurred when
the star is still on the AGB (AGB Final Thermal Pulse, AFTP), or as a
``Late'' or ``Very Late'' Thermal Pulse (LTP or VLTP, respectively),
which brings the star back to the beginning of the post-AGB track a
second time (born-again scenario). Detailed calculations
(Herwig \cite{Herwig01}) predict that the surface composition after 
an LTP and VLTP resembles the former  
intershell abundances (see Fig.~\ref{fig:LTPabundances}).
In the case of the VLTP the star has almost reached the white dwarf cooling
track when the last thermal pulse occurs. The envelope is 
then mixed downwards, and all remaining hydrogen is burnt, 
producing a few percent of nitrogen (Werner \& Herwig \cite{Werner06}). 
A corresponding abundance of nitrogen and the total absence of hydrogen
on the stellar surface would therefore be indicative for the VLTP
scenario.

\section{The Potsdam Wolf-Rayet (PoWR) model atmospheres}

Wolf-Rayet spectra form in expanding stellar atmospheres under extreme
non-LTE conditions. They often have complex ionization stratifications,
and different spectral lines form in different spatial regions. The
hydrostatic layers are often completely hidden under the optically thick
wind. Adequate model atmospheres are prerequisite to analyze Wolf-Rayet
type spectra.

The Potsdam Wolf-Rayet (PoWR) code (e.g.\ Hamann \& Gr\"afener
\cite{HamannGraefener2004}) solves the radiative transfer equation in
the co-moving frame, consistently with the equations of statistical
equilibrium. Iron line blanketing is included in the superlevel
approach. Clumping is taken into account in the approximation of
small-scale inhomogeneities, generally adopting a density contrast of
$D$=10 (Hamann \& Koesterke \cite{Hamann98}).

\begin{table}[b]
\begin{center}
\caption{Analyses of early-type [WC] stars; $\log L/L_{\odot}= 3.7$,
$M = 0.6\,M_{\odot}$, $D = 10$}
\begin{tabular}{lll|lll|lll}
\hline 
Name & Type & Type & $T_*$ & $\log R_{\rm t}$ & $v_{\infty}$ 
         & $X_{\text{He}}$ & $X_{\text{C}}$ & $X_{\text{O}}$ \\
& (1) & (2) & $[\text{kK}]$ & $[\text{R}_{\odot}]$ & $[\text{km/s}]$ &
\multicolumn{3}{c}{mass fraction}   \\
\hline 
NGC\,5189   & [WO1] & [WC2] & 144 & 0.75& 2000 & 0.62 & 0.24 & 0.14 \\
PB\,6       & [WO1] & [WC2] & 144 & 0.75& 1600 & 0.61 & 0.25 & 0.12 \\
NGC\,2452   & [WO1] & [WC2] & 141 & 0.75& 3000 & 0.64 & 0.25 & 0.11 \\
NGC\,6905   & [WO2] & [WC2] & 135 & 0.80& 2000 & 0.54 & 0.40 & 0.06: \\
NGC\,2867   & [WO2] & [WC2] & 126 & 0.95& 2500 & 0.60 & 0.25 & 0.15: \\
Hen\,2-55   & [WO3] & [WC3] & 126 & 1.00& 2200 & 0.58 & 0.30 & 0.12 \\
NGC\,6369   & [WO3] & [WC4] & 159 & 0.25& 1200 & 0.56 & 0.30 & 0.14 \\
NGC\,1501   & [WO4] & [WC4] & 141 & 0.75& 1800 & 0.55 & 0.35 & 0.10 \\
$[$S71d$]$3 &       & [WC3] & 160 & 2.50& 2000 & 0.62 & 0.26 & 0.12 \\
NGC\,7026   &       & [WC3] & 141 & 0.75& 2500 & 0.69 & 0.25 & 0.06 \\
IC\,1747    &       & [WC4] & 115 & 0.95& 2000 & 0.65 & 0.30 & 0.05 \\
\hline 
\end{tabular}

\small{
(1) new classification  (Acker \& Neiner \cite{Acker03})}
~~~(2) old classification 
\end{center}
\end{table}

\section{Spectral analysis}

Distances of Galactic PNe are generally unknown, and hence the stellar 
parameters cannot be determined on an absolute scale. We adopt a 
default luminosity of $\log L/L_\odot$ = 3.7. Fortunately expanding
stellar atmosphere follow a kind of scale invariance. The important
parameter is the ``transformed radius'' $R_{\rm t}$ as a combination of
mass-loss rate and stellar radius (cf.\ e.g.\ Hamann \& Gr\"afener
\cite{HamannGraefener2004}). The stellar mass is only of marginal
influence on the spectra and adopted to 0.6\,$M_\odot$. Hence the 
free parameters to be derived by the spectral analyses are the stellar 
temperature $T_\ast$, the transformed radius $R_{\rm t}$, the terminal 
wind velocity $v_\infty$, and, most interesting for our questions, the 
chemical composition. The results of our new analyses are compiled in 
Table\,1. In the following we briefly discuss the chemical abundances.  

\begin{figure}[t]
  \vbox{
    \hbox to \hsize{
      \begin{minipage}[b]{.5\textwidth}
	\begin{center}
        \includegraphics[width=5.5cm,height=7.2cm]{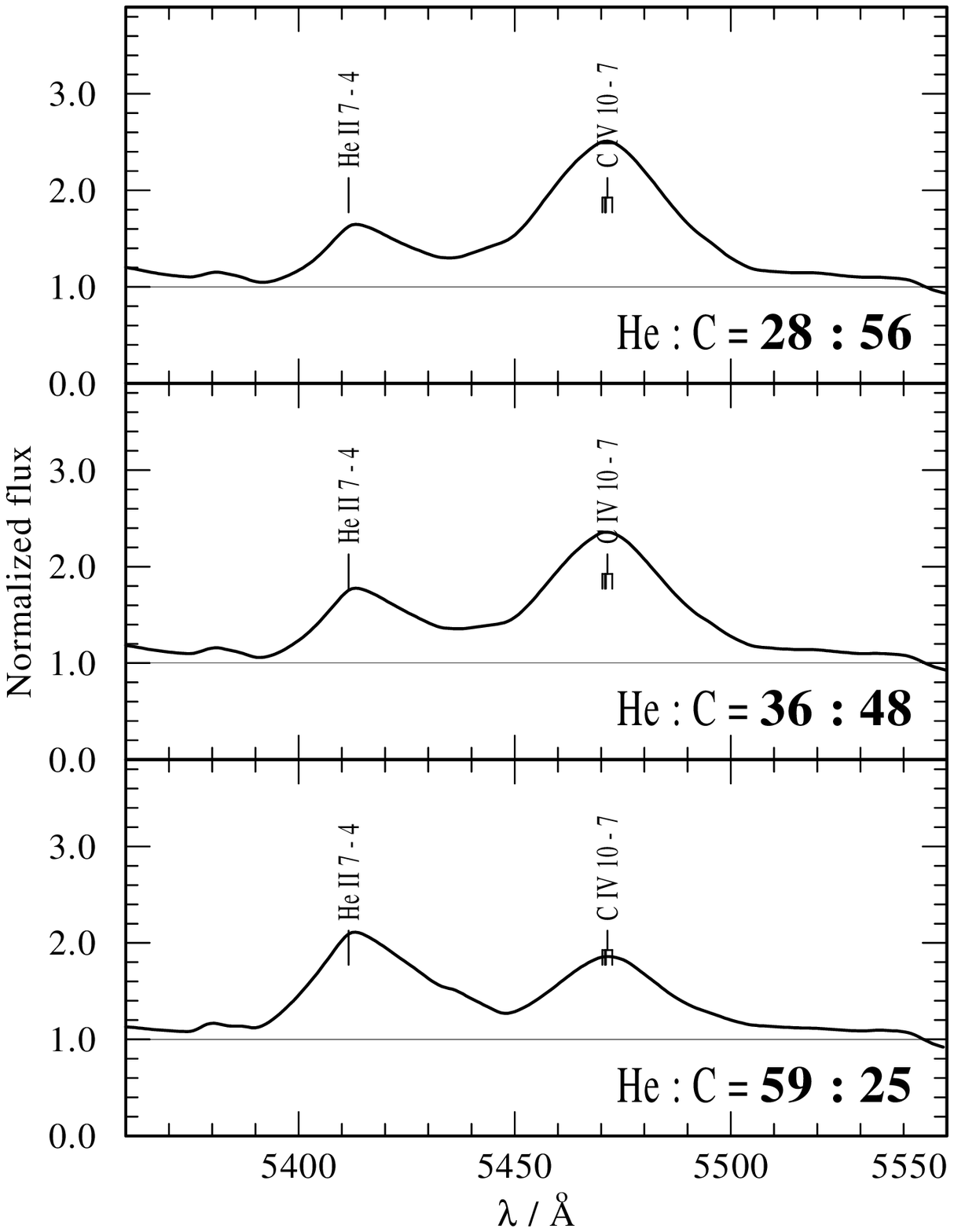}
	\end{center}
      \end{minipage}
      \hfill
      \begin{minipage}[b]{.5\textwidth}
        \center{\includegraphics[width=5.5cm,height=7.2cm]{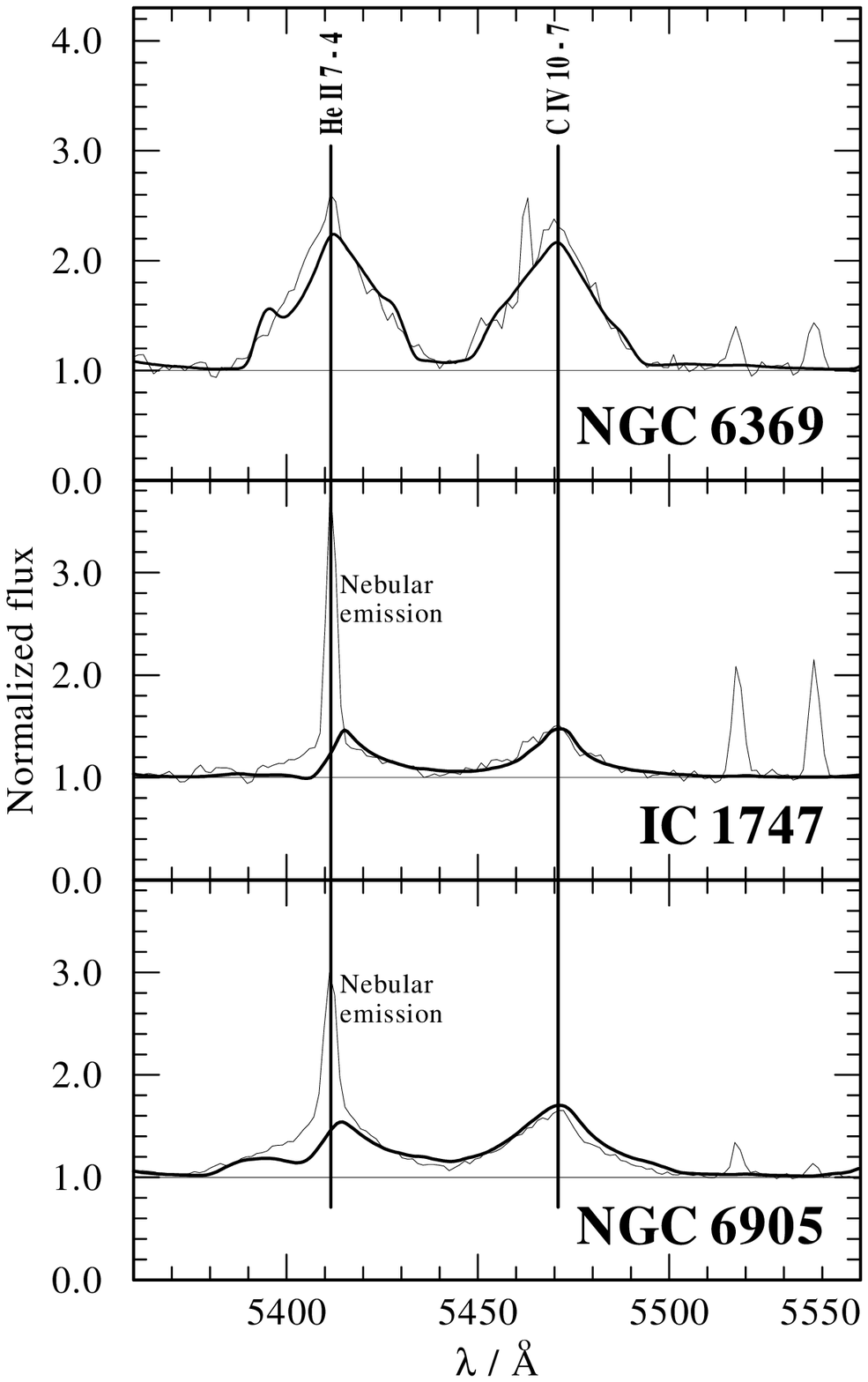}}
      \end{minipage}
    }
    \hbox to \hsize{
      \parbox[t]{.45\textwidth}{\caption{The pair of neigh\-boring
lines He\,{\sc ii}\,5412
and C\,{\sc iv}\,5470 for mod\-els with different He:C mass ratios as given in 
the plots, while all other para\-meters are the same for the whole model 
series. Roughly equal line strength is only achieved for a He:C mass 
ratio of about 2:1.
      \label{fig:hec-sequence}}}
      \hfill
      \parbox[t]{.45\textwidth}{\caption{The diagnostic line pair
He\,{\sc ii}\,5412 and C\,{\sc iv}\,5470 for
three of the program stars. The two lines are ge\-nerally observed to have
similar strength (thin lines). Fitting models have a typical He:C
mass ratio of 2:1 (thick lines). Note that the He\,{\sc ii}\,5412 is often
contaminated by a strong, narrow nebular emission.\label{fig:fit-hec}}}
    }
  }
\end{figure}

\emph {Carbon and helium.}
In our analyses we determine the He:C ratio mainly from the line pair 
He\,{\sc ii}\,5412 and C\,{\sc iv}\,5470
(Figs.~\ref{fig:hec-sequence} and \ref{fig:fit-hec}). As [WCE] have
highly ionized atmospheres, we checked the possible impact of very high
ions of carbon and oxygen which are usually neglected in the model
codes. Although the C\,{\sc iv} lines are fed mainly by recombination
from C\,{\sc v}, the inclusion of further levels above the C\,{\sc v}
singlet ground state did barely change the model spectra. Our new [WCE]
analyses confirm in all cases He:C mass ratios of about 2:1, in contrast
to the higher carbon abundance (He $\approx$ C) found in [WCL].

\begin{figure}[b]
\begin{center}
\includegraphics[width=0.9\textwidth]{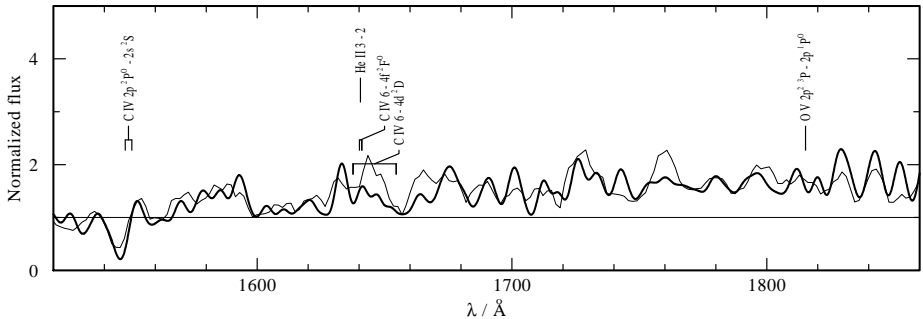}
\caption{``Iron forest'' in the late-type [WC] spectrum of BD+30$^\circ$3639 
(thin line), compared to a model (thick
line) with  $X_{\rm Fe} = 1.6\cdot10^{-3}$, $T=47\,{\rm kK}$, $\log(
R_{\rm t}/R_{\odot}) = 0.8$, $v_{\infty}=700\,$km/s.}
\label{fig:bd30IUEeisenProce}
\end{center}
\end{figure}

\emph{Nitrogen.}
As pointed out by Werner \& Herwig (\cite{Werner06}), only a VLTP can
efficiently produce nitrogen by H ingestion, leading to a N mass
fraction of a few percent. Hence the detection of nitrogen could
discriminate between the scenarios. We found indications for such
overabundance in most of our analyzed [WCE] stars (except Hen\,2-55 and
NGC\,6905), but this preliminary result must still be further
substantiated.

\emph{Hydrogen.}
In principle H abundances could discriminate between the scenarios, but
the H lines are always blended with He\,{\sc ii} and nebular emission.
Small amounts of hydrogen have been found in some [WCL] stars 
(Koesterke \cite{Koesterke01} and references therein), while some 
PG\,1159 stars -- presumably also 
evolutionary related to the [WC]-type central stars -- show even 
more hydrogen. 
In the hot [WCE] stars, a possible spectral signature from hydrogen would 
be weak. For our [WCE] program stars we can establish an upper limit 
of $5 \ldots 10$\% for the hydrogen mass fraction. 

\emph{Iron.}
A whole ``forest'' of iron lines is often visible in the UV spectra of
[WCL] central stars (see Fig.\,\ref{fig:bd30IUEeisenProce}). Our sample
of [WCE] stars, however, does not show detectable iron lines, neither in
the observed spectra nor in the models. Therefore we cannot check the
iron abundance for a possible depletion, as it was observed for one
cooler [WCE] star in the LMC, SMP\,61 (Stasinska et al.\
\cite{SMP61}).

\begin{figure}[t]
\begin{center}
\includegraphics[width=\textwidth]{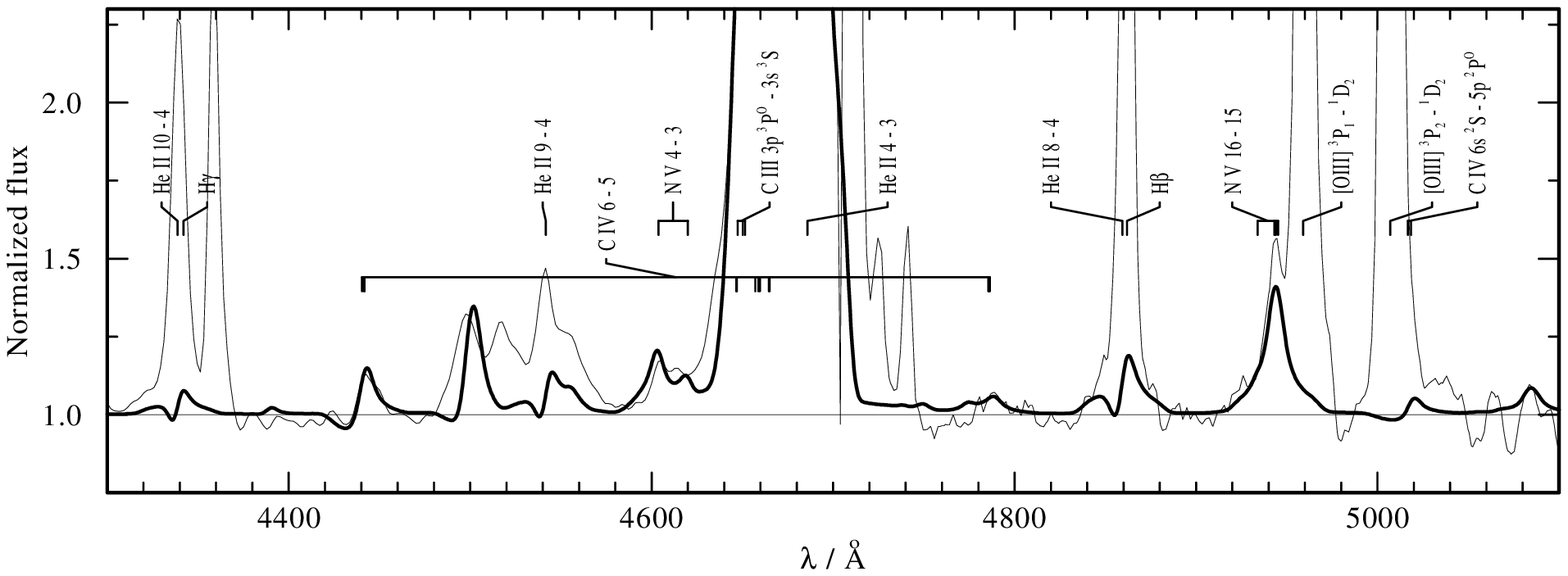}
\caption{PB\,6: Model (thick line) with $X_{\rm N}$\,=\,1.5\%
  and observation (thin line)}
\end{center}
\end{figure}

\section{Conclusions}

The comparison between spectral analyses and evolutionary predictions 
are still not conclusive. The low carbon abundance of [WCE] stars is 
contrasting to the [WCL] stars, and pointing to an AFTP origin. The small 
amount of H found in some [WCL] is explained from LTP models, 
while the possible N overabundance is rather predicted by a ``Very Late 
Thermal Pulse (VLTP)''.

\begin{acknowledgments}
We thank Adriane Liermann and Andreas
Barniske for their help. This work was supported by the
Bundesministerium f\"ur Bildung und Forschung (BMBF) under grant 05AVIPB/1.
\end{acknowledgments}

\end{document}